\documentclass[pdftex,a4paper,twoside,british,galaxies,article,submit,moreauthors,pdftex,10pt,a4paper]{mdpi}
\pdfoutput=1
\preto{\abstractkeywords}{\nolinenumbers}
\usepackage[T1]{fontenc}
\usepackage{fancyhdr}
\usepackage{amsmath}
\usepackage{amssymb}
\usepackage{graphicx}
\makeatletter
\Title{Post-AGB discs from common-envelope evolution}
\Author{Robert~G.~Izzard$^{1,2}$\orcidA{0000-0003-0378-4843} and Adam~S.~Jermyn$^{2}$\orcidB{0000-0001-5048-9973}}
\AuthorNames{Robert G. Izzard and Adam S.~Jermyn}
\address{$^{1}$\quad{}Department of Physics, University of Surrey, Guildford,
Surrey, GU2 7XH, United Kingdom. r.izzard@surrey.ac.uk\\
$^{2}$\quad{}Institute of Astronomy, Madingley Road, Cambridge,
CB3 0HA, United Kingdom. adamjermyn@gmail.com }
\corres{Correspondence: r.izzard@surrey.ac.uk}
\abstract{Post-asymptotic giant branch (post-AGB) stars with discs are all
binaries. Many of these binaries have orbital periods between $100$
and $1000\,\mathrm{days}$ so cannot have avoided mass transfer between
the AGB star and its companion, likely through a common-envelope type
interaction. We report on preliminary results of our project to model
circumbinary discs around post-AGB stars using our binary population
synthesis code \emph{binary\_c}. We combine a simple analytic thin-disc
model with binary stellar evolution to estimate the impact of the
disc on the binary, and vice versa, fast enough that we can model
stellar population and hence explore the rather uncertain parameter
space involved with disc formation. We find that, provided the discs
form with sufficient mass and angular momentum, and have an inner
edge that is relatively close to the binary, they can both prolong
the life of their parent post-AGB star and pump the eccentricity of
orbits of their inner binaries.}
\pdfpageheight\paperheight
\pdfpagewidth\paperwidth

\let\jnl@style=\rmfamily 
\def\ref@jnl#1{{\jnl@style#1}}% 
\providecommand\aj{\ref@jnl{AJ}}%        % Astronomical Journal 
\providecommand\araa{\ref@jnl{ARA\&A}}%  % Annual Review of Astron and Astrophys 
\providecommand\apj{\ref@jnl{ApJ}}%    % Astrophysical Journal ++
\providecommand\apjl{\ref@jnl{ApJL}}     % Astrophysical Journal, Letters 
\providecommand\apjs{\ref@jnl{ApJS}}%    % Astrophysical Journal, Supplement 
\providecommand\ao{\ref@jnl{ApOpt}}%   % Applied Optics ++
\providecommand\apss{\ref@jnl{Ap\&SS}}%  % Astrophysics and Space Science 
\providecommand\aap{\ref@jnl{A\&A}}%     % Astronomy and Astrophysics 
\providecommand\aapr{\ref@jnl{A\&A~Rv}}%  % Astronomy and Astrophysics Reviews 
\providecommand\aaps{\ref@jnl{A\&AS}}%    % Astronomy and Astrophysics, Supplement 
\providecommand\azh{\ref@jnl{AZh}}%       % Astronomicheskii Zhurnal 
\providecommand\baas{\ref@jnl{BAAS}}%     % Bulletin of the AAS 
\providecommand\icarus{\ref@jnl{Icarus}}% % Icarus
\providecommand\jrasc{\ref@jnl{JRASC}}%   % Journal of the RAS of Canada 
\providecommand\memras{\ref@jnl{MmRAS}}%  % Memoirs of the RAS 
\providecommand\mnras{\ref@jnl{MNRAS}}%   % Monthly Notices of the RAS 
\providecommand\pra{\ref@jnl{PhRvA}}% % Physical Review A: General Physics ++
\providecommand\prb{\ref@jnl{PhRvB}}% % Physical Review B: Solid State ++
\providecommand\prc{\ref@jnl{PhRvC}}% % Physical Review C ++
\providecommand\prd{\ref@jnl{PhRvD}}% % Physical Review D ++
\providecommand\pre{\ref@jnl{PhRvE}}% % Physical Review E ++
\providecommand\prl{\ref@jnl{PhRvL}}% % Physical Review Letters 
\providecommand\pasp{\ref@jnl{PASP}}%     % Publications of the ASP 
\providecommand\pasj{\ref@jnl{PASJ}}%     % Publications of the ASJ 
\providecommand\qjras{\ref@jnl{QJRAS}}%   % Quarterly Journal of the RAS 
\providecommand\skytel{\ref@jnl{S\&T}}%   % Sky and Telescope 
\providecommand\solphys{\ref@jnl{SoPh}}% % Solar Physics 
\providecommand\sovast{\ref@jnl{Soviet~Ast.}}% % Soviet Astronomy 
\providecommand\ssr{\ref@jnl{SSRv}}% % Space Science Reviews 
\providecommand\zap{\ref@jnl{ZA}}%       % Zeitschrift fuer Astrophysik 
\providecommand\nat{\ref@jnl{Nature}}%  % Nature 
\providecommand\iaucirc{\ref@jnl{IAUC}}% % IAU Cirulars 
\providecommand\aplett{\ref@jnl{Astrophys.~Lett.}}%  % Astrophysics Letters 
\providecommand\apspr{\ref@jnl{Astrophys.~Space~Phys.~Res.}}% % Astrophysics Space Physics Research 
\providecommand\bain{\ref@jnl{BAN}}% % Bulletin Astronomical Institute of the Netherlands 
\providecommand\fcp{\ref@jnl{FCPh}}%   % Fundamental Cosmic Physics 
\providecommand\gca{\ref@jnl{GeoCoA}}% % Geochimica Cosmochimica Acta 
\providecommand\grl{\ref@jnl{Geophys.~Res.~Lett.}}%  % Geophysics Research Letters 
\providecommand\jcp{\ref@jnl{JChPh}}%     % Journal of Chemical Physics 
\providecommand\jgr{\ref@jnl{J.~Geophys.~Res.}}%     % Journal of Geophysics Research 
\providecommand\jqsrt{\ref@jnl{JQSRT}}%   % Journal of Quantitiative Spectroscopy and Radiative Trasfer 
\providecommand\memsai{\ref@jnl{MmSAI}}% % Mem. Societa Astronomica Italiana 
\providecommand\nphysa{\ref@jnl{NuPhA}}%     % Nuclear Physics A 
\providecommand\physrep{\ref@jnl{PhR}}%       % Physics Reports 
\providecommand\physscr{\ref@jnl{PhyS}}%        % Physica Scripta 
\providecommand\planss{\ref@jnl{Planet.~Space~Sci.}}%  % Planetary Space Science 
\providecommand\procspie{\ref@jnl{Proc.~SPIE}}%      % Proceedings of the SPIE 

\providecommand\actaa{\ref@jnl{AcA}}%  % Acta Astronomica
\providecommand\caa{\ref@jnl{ChA\&A}}%  % Chinese Astronomy and Astrophysics
\providecommand\cjaa{\ref@jnl{ChJA\&A}}%  % Chinese Journal of Astronomy and Astrophysics
\providecommand\jcap{\ref@jnl{JCAP}}%  % Journal of Cosmology and Astroparticle Physics
\providecommand\na{\ref@jnl{NewA}}%  % New Astronomy
\providecommand\nar{\ref@jnl{NewAR}}%  % New Astronomy Review
\providecommand\pasa{\ref@jnl{PASA}}%  % Publications of the Astron. Soc. of Australia
\providecommand\rmxaa{\ref@jnl{RMxAA}}%  % Revista Mexicana de Astronomia y Astrofisica

\providecommand\maps{\ref@jnl{M\&PS}}% Meteoritics and Planetary Science
\providecommand\aas{\ref@jnl{AAS Meeting Abstracts}}% American Astronomical Society Meeting Abstracts
\providecommand\dps{\ref@jnl{AAS/DPS Meeting Abstracts}}% American Astronomical Society/Division for Planetary Sciences Meeting Abstracts

\providecommand\cac{\ref@jnl{Computational Astrophysics and Cosmology}}

\firstpage{1}
\articlenumber{x}
\doinum{10.3390/------}
\pubvolume{xx}
\pubyear{2018}
\copyrightyear{2018}
\history{Received: date; Accepted: date; Published: date}

\pdfoutput=1 

\usepackage{xcolor}
\newcommand{\change}[1]{#1}

\makeatother

\usepackage{babel}
\begin{document}
\maketitle

\section{Introduction\label{sec:Introduction}}

About half the stars exceeding the mass of the Sun are binaries, and
of these many interact during their lifetime \cite{2017PASA...34....1D}.
Many key astrophysical phenomena occur in binary systems, such as
thermonuclear novae, X-ray bursts, type Ia supernovae and merging
compact objects as detected by gravitational waves. Despite their
importance to fundamental astrophysics, many aspects of binary evolution
remain poorly understood. 

The evolution of binary stars differs from that of single stars mostly
because of mass transfer. In binaries wide enough that one star becomes
a giant, yet short enough that the giant cannot fit inside the binary,
mass transfer begins when \change{the} radius of the giant star
exceeds its Roche radius. Because the giant is convective, and likely
significantly more massive than its companion, mass transfer is unstable
and accelerates. This leads to a common envelope forming around the
stars. Drag between the stars and the envelope cause the orbit to
decay and, if enough energy is transferred to the envelope, the envelope
is ejected \cite{2013A&ARv..21...59I}. Stars on the giant branch
or asymptotic giant branch (GB or AGB stars) are stripped almost down
to their cores, leaving a post-(A)GB binary. 

During the common envelope process, tides are expected to be highly
efficient and hence the binary that emerges is expected to have little,
if any, eccentricity ($e\approx0$). Yet, post-AGB binaries are often
highly eccentric, up to $e=0.6$. The source of this eccentricity
is unknown but is probably related to a similar phenomena observed
in the barium stars which are thought to have involved mass transfer
from an AGB star \cite{1998A&A...332..877J,2000MNRAS.316..689K}. 

Many post-AGB stars have discs, and all those with discs are in binary
systems \cite{2013A&A...557A.104B}. Investigations into the link
between these discs and their stellar systems' peculiar eccentricity,
based on eccentricity pumping by Lindblad resonances, suggest that
the discs can cause the post-AGB systems' eccentricity if they are
sufficiently massive and live for long enough \cite{2013A&A...551A..50D}.
Recent ALMA observations, e.g.~of IRAS 08544-4431, show that post-AGB
discs are mostly Keplerian, have masses of about $10^{-2}\mathrm{\,M_{\odot}}$,
outer diameters of $10^{16}\,\mathrm{cm}\approx10^{5}\mathrm{\,R_{\odot}}$,
angular momenta similar to their parent binary systems (around $10^{52}\,\mathrm{g}\,\mathrm{cm}^{2}\,\mathrm{s}^{-1}$),
and both slow mass loss from the outer part of the disc and inflow
at its inner edge at a rate of about $10^{-7}\mathrm{\,M_{\odot}}\,\mathrm{year}^{-1}$
\cite{2018A&A...614A..58B}.

In this work we combine a fast, analytic model of circumbinary discs
with a synthetic binary stellar-evolution code to estimate the number
of post-(A)GB discs and their properties. We include mass loss from
the disc caused by illumination from the central star, ram-stripping
by the interstellar medium, and include a viscous-timescale flow onto
the central binary. The disc extracts angular momentum from its central
binary star system. Resonances excited in the disc pump the central
binary's eccentricity. While our results are preliminary, they show
that eccentricity pumping is efficient in some systems, and the discs
may live a considerable time.

\section{Circumbinary disc model\label{sec:Circumbinary-disc-model}}

We assume that circular, Keplerian discs form when a common envelope
is ejected and some small fraction of the envelope mass, $f_{M}\ll1$,
is left behind as a disc containing a fraction, $f_{J}\ll1$, of the
envelope's angular momentum. The disc thus has mass $M_{\mathrm{disc}}$
and angular momentum $J_{\mathrm{disc}}$. We assume our discs are
thin such that $H/R<1$, where $H(R)$ is the scale height at a radius
$R$, and have viscous timescales that are short compared to their
lifetimes such that they spread instantaneously \cite{1974MNRAS.168..603L}.
Given that observed discs are mostly cool and neutral (i.e.~$T\lesssim1000\,\mathrm{K}$
in most of the disc), we fix the opacity to $\kappa=0.01\,\mathrm{m}^{2}\,\mathrm{g}^{-1}$
and assume an $\alpha$-viscosity model with $\alpha=10^{-3}$.

Given the above simplifications, we write the heat-balance equation
in the disc as,
\begin{alignat}{1}
\sigma T^{4} & ={\cal A}+{\cal B}\left(1+{\cal C}\right)\,,\label{eq:disc-heat-balance}
\end{alignat}
where $T=T(R)$ is the temperature in the disc mid-plane (cf.~\cite{2013ApJ...764..169P}
where these terms are derived in detail). The $\sigma T^{4}$ term
is the heating by the post-(A)GB star, which is balanced in equilibrium
by visocus heating (${\cal A}$) and re-radiation (${\cal B}$ and
${\cal C}$). We neglect the impact of mass changes on the heat balance
because observed discs have low mass-loss rates ($\lesssim10^{-7}\mathrm{\,M_{\odot}}\,\mathrm{year}^{-1}$).
The terms ${\cal A}$, ${\cal B}$ and ${\cal C}$ are functions of
radius, such that we can rewrite Eq.~\ref{eq:disc-heat-balance}
as, 
\begin{eqnarray}
T^{4} & = & a\Sigma^{2}TR^{-3/2}+bT^{1/2}R^{-3/2}+cR^{-3},\label{eq:heat-balance2}
\end{eqnarray}
where $\Sigma=\Sigma(R)$ is the surface density at radius $R$. At
a given radius $R$ we then choose the largest term in the right hand
side of Eq.~\ref{eq:heat-balance2} and set the other terms to zero,
allowing a simple solution for $T(R)$ everywhere in the disc (cf.\textbf{~}\cite{2009ApJ...700.1952H}).
Errors in $T^{4}$ are up to a factor of 3, hence errors in $T$ are
at most $3^{1/4}\approx30\%$ and typically much less, which is good
enough for our purposes.

\change{To verify that we can neglect mass changes, note that the
temperature, $T_{\mathrm{acc}}$, associated with accretion at a rate
$\dot{M}$ at radius $R$ and binary mass $M_{\mathrm{binary}}$ \cite{2013ApJ...764..169P},

\begin{eqnarray}
\sigma T_{{\rm acc}}^{4} & = & \frac{GM_{{\rm binary}}\dot{M}}{2\pi R^{3}}\,,\label{eq:accretion-heat}
\end{eqnarray}
where the gravitational mass is dominated by the binary. Evaluating
this yields,

\begin{eqnarray}
T_{{\rm acc}}\left(R\right) & = & 290{\rm K}\left(\frac{\dot{M}}{10^{-7}\mathrm{M}_{\odot}\,{\rm year}^{-1}}\right)^{1/4}\left(\frac{M_{{\rm binary}}}{\mathrm{M}_{\odot}}\right)^{1/4}\left(\frac{R}{10^{2}\mathrm{R}_{\odot}}\right)^{-3/4}\,.\label{eq:accretion-temperature}
\end{eqnarray}
At the inner edges of our discs $R\approx10^{2}\mathrm{R}_{\odot}$
so $T_{{\rm acc}}$ is a factor of several smaller than the typical
$1000{\rm K}$ temperatures we find. The outer edges, at $R\approx10^{5}\mathrm{R}_{\odot}$,
have $T_{{\rm acc}}\approx2\,\mathrm{K}$ which both cooler than the
disc and the Cosmic Microwave background so this term can be safely
ignored.}

We next scale the density and outer radius so that the integrals of
mass and angular momentum throughout the disc match $M_{\mathrm{disc}}$
and $J_{\mathrm{disc}}$ respectively, while the inner radius $R_{\mathrm{in}}$
is fixed by the torque on the disc caused by the inner binary which
we take from \cite{2002ApJ...567L...9A} with a multiplier of $10^{-3}$.
Thus, given three constraints -- $M_{\mathrm{disc}}$, $J_{\mathrm{disc}}$
and the inner binary torque -- we know $T(R)$ throughout the disc.
From this we construct any other required physical quantities, and
can integrate these to calculate, for example, the luminosity of the
disc. With our chosen torque prescription, our discs' inner radii
are typically around twice the orbital separation, as assumed in other
works (e.g.~\cite{2016ApJ...830....8R}).

We treat mass loss from the circumbinary disc as a slow phenomenon.
At the inner edge of the disc we include mass inflow onto the central
binary at the local viscous rate. At the outer edge we strip material
when its pressure is less than that of the interstellar medium, assumed
to be $3000\text{K}/k_{\mathrm{B}}$ where $k_{\mathrm{B}}$ is the
Boltzmann constant. Irradiation by the post-(A)GB star, particularly
in X-rays, causes mass loss and we model this with the prescription
of \cite{2012MNRAS.422.1880O}. During most of the disc's lifetime,
the viscous inflow is most important, but X-ray losses dominate when
the star transitions to become a hot, young white dwarf and these
losses quickly evaporate the disc. Our neglect of mass loss in Eq.~\ref{eq:disc-heat-balance}
is incorrect in this brief phase but because the disc is terminated
very rapidly, on timescales of years, such systems will be rarely
observed.

The evolution of the binary stars is calculated using \emph{binary\_c}
\cite{2004MNRAS.350..407I,2006A&A...460..565I,2009A&A...508.1359I,2018MNRAS.473.2984I}.
Common envelopes are ejected with the formalism of \cite{2002MNRAS_329_897H}
using an efficiency $\alpha_{\mathrm{CE}}=1$ \change{to match observed
post-AGB systems with periods between $100$ and about $1000\,\mathrm{d}$
which are observed to have circumbinary discs}. The envelope binding
energy parameter, $\lambda_{\mathrm{CE}}$, is fitted to the models
of \cite{2001A&A...369..170T}, and $10\%$ of the envelope's recombination
energy is used to aid ejection. Because typically $\lambda_{\mathrm{CE}}\gg1$
during the AGB, such envelopes are nearly unbound and common envelope
ejection is efficient with only modest orbital shrinkage.First giant
branch stars have $\lambda\lesssim1$ because they are more tightly
bound, \change{so their orbits} shrink significantly. We also assume
that stars exit the common envelope with a small eccentricity, $e=10^{-5}$,
to which we apply Lindblad resonance pumping \cite{2013A&A...551A..50D}.
\change{If $\alpha_{\mathrm{CE}}$ is less than 1.0, as suggested
by e.g.~\cite{2010A&A...520A..86Z,2011MNRAS.411.2277D,2012MNRAS.419..287D},
then more recombination energy can be included to prevent orbital
shrinkage. We do not pretend to better understand common envelope
evolution with our simple model.}

At the end of the common envelope phase our treatment differs from
\cite{2002MNRAS_329_897H} in that we keep a thin envelope on the
(A)GB star such that it just fills its Roche lobe when the envelope
is ejected. Typically this is a $10^{-2}-10^{-3}\mathrm{\,M_{\odot}}$
hydrogen-rich envelope which keeps the star cool ($\sim5000\,\mathrm{K}$)
during the post-(A)GB phase relative to the white dwarf ($\gtrsim10^{4}\,\mathrm{K}$)
it will become. The star continues its nuclear burning which reduces
the mass of the envelope. Accretion from the circumbinary disc replenishes
the shell and extends the lifetime of the post-(A)GB phase, as we
show in the following section. 

\section{Example system\label{sec:Example-system}}

As an example binary star system we choose an initial primary mass
$M_{1}=1.5\mathrm{\,M_{\odot}}$, initial secondary mass $M_{2}=0.9\mathrm{\,M_{\odot}}$,
initial separation $a=800\mathrm{\,R_{\odot}}$ and metallicity $Z=0.02$.
The separation is chosen such that Roche-lobe overflow is initiated
just after the primary starts thermally pulsing on the AGB. Common
envelope evolution follows with $\alpha_{\mathrm{CE}}\lambda_{\mathrm{CE}}=2.1$
so the orbit shrinks to $195\mathrm{\,R_{\odot}}$. The primary is
then a $0.577\mathrm{\,M_{\odot}}$ post-AGB star, with an envelope
mass of $7\times10^{-3}\mathrm{\,M_{\odot}}$ and a $0.95\mathrm{\,M_{\odot}}$
main-sequence dwarf companion. \change{The $0.05\mathrm{\,M_{\odot}}$
accreted on to the secondary is from the wind of the AGB star prior
to common envelope evolution.} The orbital period is then about $255\,\mathrm{days}$,
typical of post-AGB binaries with circumbinary discs. \change{We
model circular orbits but our model is equally applicable to initially
mildly eccentric binaries. Tides are expected to be efficient as the
primary ascends the AGB and will quickly circularize the system. In
the following discussion, and the figures, times are measured from
the moment the common envelope is ejected.}

\begin{figure}
\begin{centering}
\includegraphics[scale=0.6]{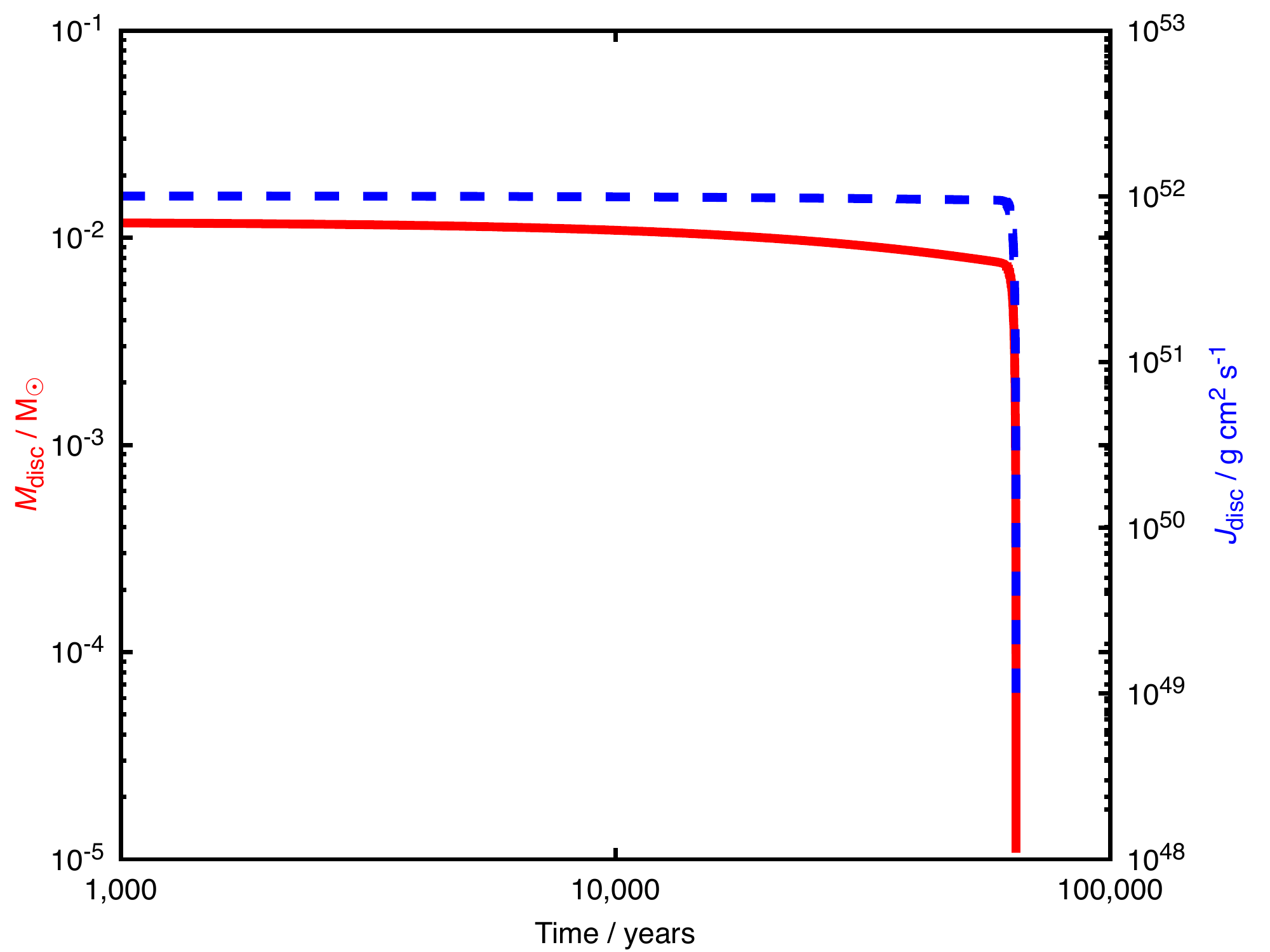}
\par\end{centering}
\caption{\label{fig:Fig1-MJ}Mass (left axis, red solid line) and angular momentum
(right axis, blue dashed line) evolution in our example circumbinary
disc system. During most of the evolution of the disc, its mass changes
slowly ($\sim10^{-8}\mathrm{\,M_{\odot}}\,\mathrm{year}^{-1}$) because
of viscous flow through its inner edge onto the inner binary system.
As the post-AGB star in the binary heats up, its X-ray flux increases
until it drives sufficient wind that the disc is quickly evaporated
after about $6.5\times10^{4}\,\mathrm{years}$ \change{where a time
of $0\,\mathrm{years}$ is when the common envelope is ejected}.}

\end{figure}

A circumbinary disc is formed with $f_{M}=0.02$ and $f_{J}=0.107$,
giving $M_{\mathrm{disc}}=0.012\mathrm{\,M_{\odot}}$ and $J_{\mathrm{disc}}=10^{52}\,\mathrm{g}\,\mathrm{cm}^{2}\,\mathrm{s}^{-1}$.
Both $f_{M}$ and $f_{J}$ are chosen to give us a disc with mass
and angular momentum similar that of IRAS 08544-4431 \cite{2018A&A...614A..58B}
and other post-AGB systems with circumbinary discs (e.g.~the Red
Rectangle). As the disc evolves, it feeds off the angular momentum
of the inner binary, but the total angular momentum gained during
its lifetime is small. Mass flows through the inner edge onto the
binary at between $10^{-7}$ (initially) and $10^{-8}\mathrm{\,M_{\odot}}\,\mathrm{year}^{-1}$,
although this does not significantly alter the evolution of the disc.
Rather, X-ray driven mass loss, caused by the post-AGB star increasing
in temperature at approximately constant luminosity, leads to sudden
termination of the disc at $6.5\times10^{4}\,\mathrm{years}$, as
shown in Fig.~\ref{fig:Fig1-MJ}.

\begin{figure}
\begin{centering}
\includegraphics[scale=0.6]{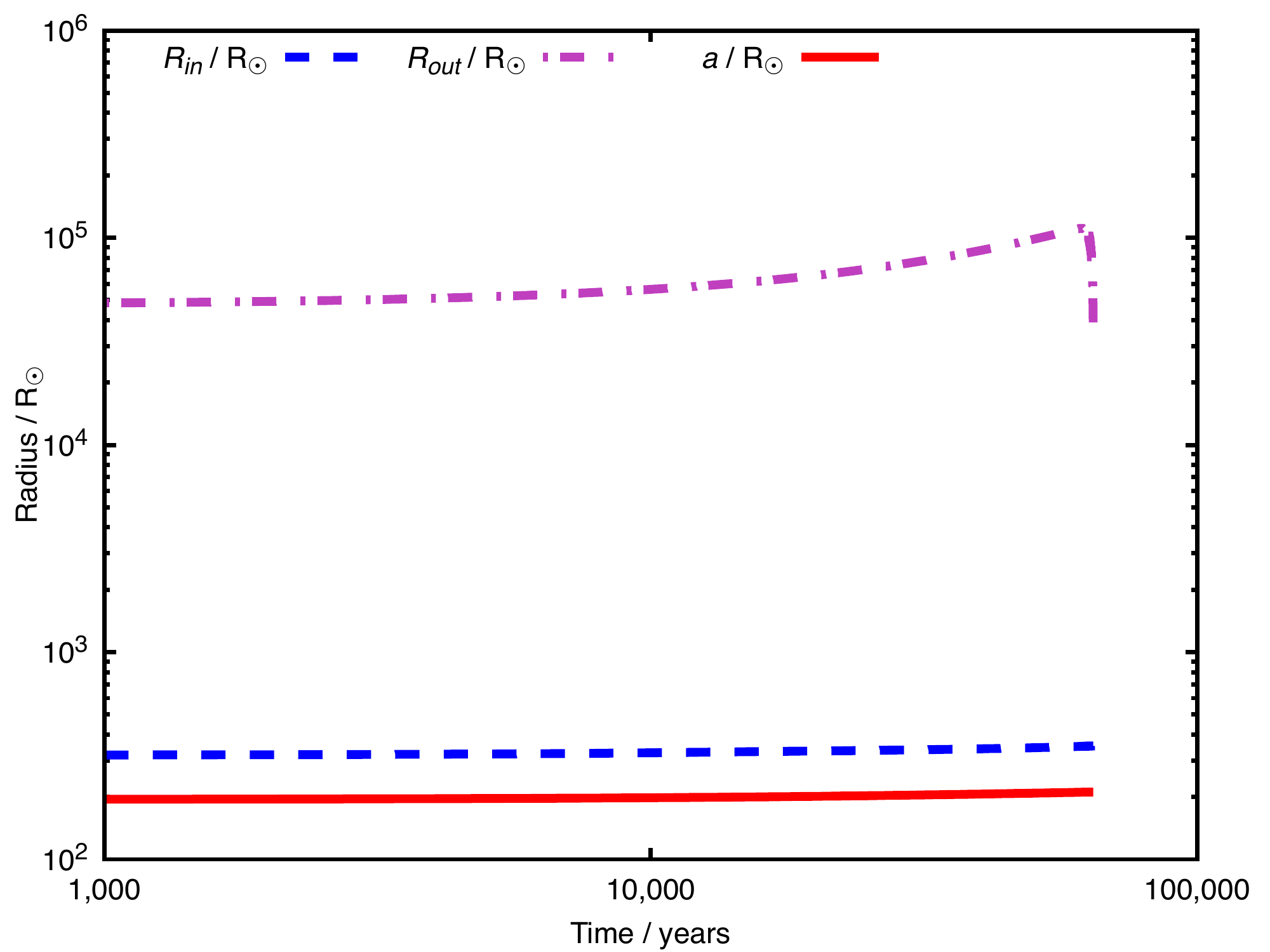}
\par\end{centering}
\caption{\label{fig:Fig2-R}Evolution of the inner and outer radii, $R_{\mathrm{in}}$
and $R_{\mathrm{out}}$ respectively, and the binary orbital separation
$a$, in our example circumbinary disc system. \change{Time is measured
from the moment of common envelope ejection.}}
\end{figure}

The inner edge of the disc is at $317\mathrm{\,R_{\odot}}$, well
outside the inner binary orbit, while the outer edge is at $4.7\times10^{4}\mathrm{\,R_{\odot}}$.
The former is set by the applied torque, while the latter is set by
the disc angular momentum. Until the disc is evaporated, neither the
inner nor outer radius changes significantly. Fig.~\ref{fig:Fig2-R}
shows the evolution of said radii. 
 The eccentricity of the inner binary system is pumped to about $0.25$
by the time the disc is evaporated. This is similar to the eccentricity
observed in post-AGB systems, and it is certainly non-zero.

Mass accreted onto the post-AGB star from the inner edge of the disc
replenishes its hydrogen-rich envelope, thus cooling the star and
prolonging its lifetime. Because our post-AGB star has a core mass
of only $0.57\mathrm{\,M_{\odot}}$, its nuclear burning rate is similar
to the disc's viscous accretion rate. The stellar wind mass loss rate
is less than $10^{-11}\mathrm{\,M_{\odot}\,\mathrm{year}^{-1}}$ because
we apply the rate of \cite{1993ApJ...413..641V} in our ignorance
of the mechanism of post-AGB wind loss. To test how long accretion
extends the post-AGB, we evolved an identical example system but with
the inner-edge viscous inflow disabled. Fig.~3 shows that the post-AGB
star in the system with accretion lives for an extra $3\times10^{4}\,\mathrm{years}$,
an approximate doubling of the its post-AGB lifetime. \change{The
extra lifetime of such systems may explain why they do not show residual
nebulosity from envelope ejection. Observed planetary nebulae, which
may be ejected common envelopes, have dynamical timescales of about
$10^{4}\,\mathrm{years}$. By the time the post-AGB star is hot enough
to ionize such envelopes they are likely too diffuse to be observed
as planetary nebulae \cite{Keller_PNe_in_prep}.}
\begin{figure}
\begin{centering}
\includegraphics[scale=0.6]{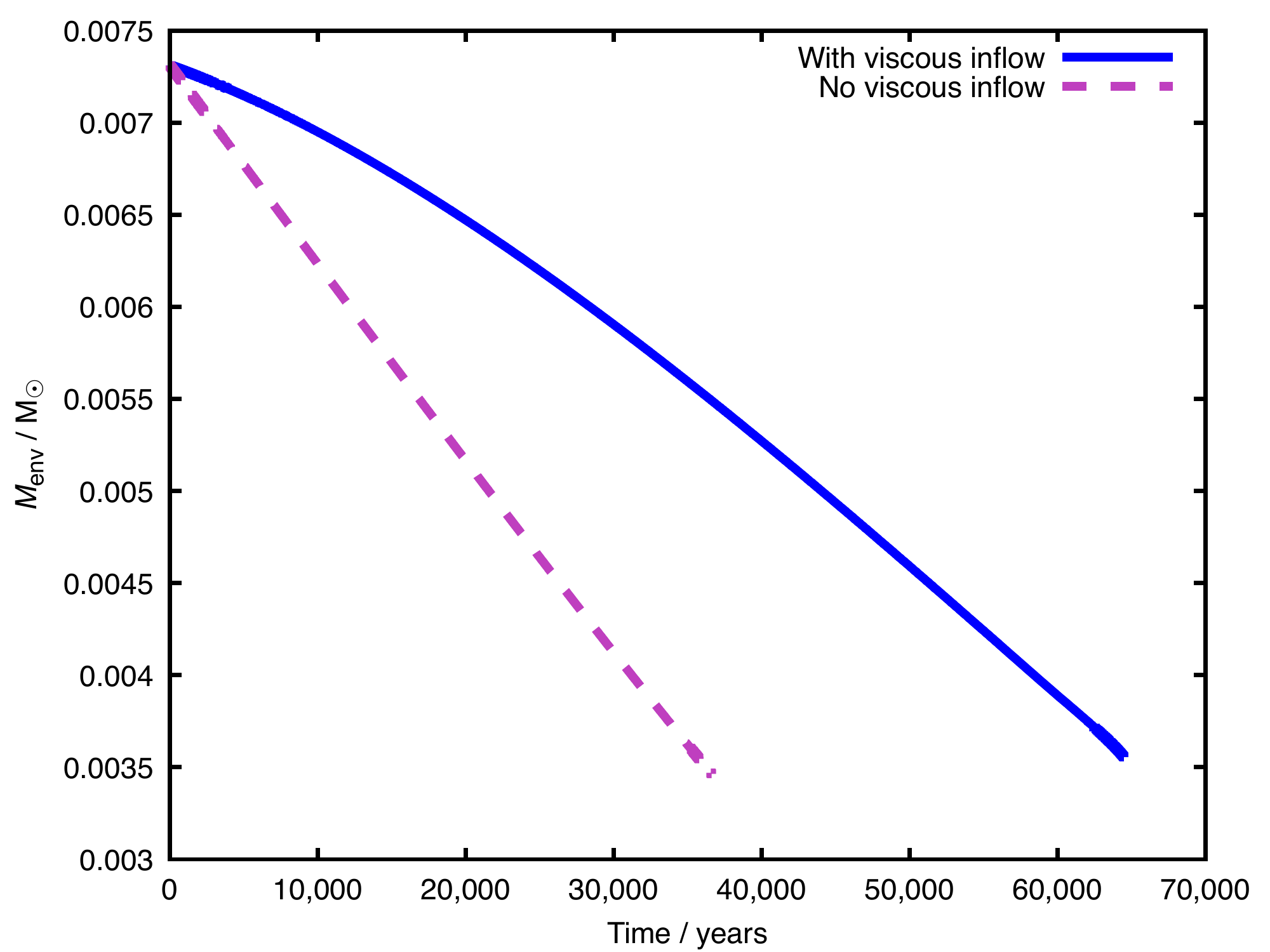}
\par\end{centering}
\caption{\label{fig:Fig1-MJ-1}Post-AGB envelope mass vs\change{.} time \change{since
common envelope ejection}. The blue, solid line is our example system
with mass inflow from the inner edge onto the inner binary, while
the magenta, dashed line is without. Mass flowing onto the post-AGB
star replenishes its hydrogen envelope thus, for a while, keeps it
relatively cool, limits its X-ray flux and prevents it from evaporating
the disc. In this case, its lifetime is extended from about $35,000\,\mathrm{y}$
to $65,000\,\mathrm{years}$. In our example system, post-AGB wind
mass loss is negligible.}
\end{figure}

\section{Stellar populations and improved physics\label{sec:Stellar-populations}}

Our model is simple yet it contains the essential physics of circumbinary
discs around post-(A)GB stars. It is also fast enough that we can
evolve a typical binary system containing a circumbinary disc in just
a few seconds of CPU time. Speed is a an essential requirement of
binary population synthesis studies because the parameter space is
large. We can explore the consequence of the initial mass and angular
momentum of our discs through the input parameters $f_{M}$ and $f_{J}$,
and we can estimate the effect of changing uncertain input physics,
e.g. a stronger or weaker X-ray wind or binary torque.

We can also model post-first giant branch (helium core) systems and
post-early-AGB systems. From an observational point of view, these
differ in that their stellar evolution is truncated at an earlier
stage than in post-thermally-pulsing-AGB, hence they are dimmer. These
systems also overflow their Roche lobes at shorter periods \change{and
have more bound envelopes}, so their orbits and resultant discs are
more compact. Our assumption of constant opacity likely breaks at
this point, although our assumed instantaneous viscous spreading of
the disc is certainly valid. We are working on improving the model
to take this into account.

Our discs are low in mass relative to their stars and we never put
more than 10\% of the common envelope mass in the disc. The example
system we report in section~\ref{sec:Example-system} has a Toomre
$Q$ parameter of about $10^{8}$ \cite{1964ApJ...139.1217T} so is
not gravitationally unstable. However, this does not preclude the
formation of rocks in the disc, after all we know the discs contain
small grains which emit in the infra red. In our models we expect
a small number of discs to form in systems that exit the red giant
branch just before helium ignition. These systems will contain sdB
stars rather than white dwarfs, thus stay relatively cool. The relatively
low X-ray flux from sdB stars means their discs are not evaporated
quickly, at least not by means modelled here, so long-lived discs
and the formation of so-called debris, i.e.~rocks, in them discs
seems quite likely.

The formation of second-generation planets in our discs seems not
to be favoured. Our discs live for less than $10^{5}\,\mathrm{years}$,
too short for planet formation in the canonical sense, and our discs
are quite warm, hotter than $1000\,\mathrm{K}$, near the inner edge
where they are densest. That said, circumbinary discs have material
concentrated in their orbital plane of the system for some time, so
if even minor planets survive the common-envelope phase, they could
accrete material from the disc. The consequences for the disc may
be that it does not survive but this is currently beyond the scope
of our model.

Our discs do succeed in pumping the eccentricity of their inner binary,
in the case of our example system up to 0.25. In part this is because
our discs are, by design, quite massive (about $0.01\mathrm{\,M_{\odot}}$),
but this is to match observed discs such as IRAS 08544-4431, so is
reasonable. Circumbinary discs seem to be good candidates to explain
the eccentricities of at least some post-mass-transfer objects such
as barium stars. We have not yet tested wind mass transfer as a mechanism
for disc formation, but this is an obvious extension to our work and
likely contributes to the observed population of post-(A)GB binaries.

\authorcontributions{Conceptualization, data curation, funding acquisition, project administration,
resources, software, supervision and writing -- original draft; Robert~G.
Izzard. Formal analysis, investigation, methodology, validation, visualization
and writing -- review and editing; Robert~G. Izzard and Adam~S.
Jermyn.}

\acknowledgments{RGI thanks the STFC for funding his Rutherford fellowship under grant
ST/L003910/1, Churchill College, Cambridge for his fellowship and
access to their library. Hello to Jason Isaacs. ASJ thanks the UK
Marshall Commission for financial support. }

\conflictsofinterest{The authors declare no conflict of interest.}

\funding{This research was funded by the Science and Technology Facilities
Council under grant number ST/L003910/1. }

%\externalbibliography{yes}\bibliography{/home/izzard/svn/tex/references}

\end{document}